\documentclass[a4paper,10pt,twocolumn]{article}

\usepackage[margin=0.8in]{geometry}
\usepackage{mathptmx}
\usepackage{helvet}

\usepackage{amsmath,amssymb}
\usepackage{graphicx}
\usepackage{titlesec}
\usepackage{caption}
\usepackage{multicol}
\usepackage{booktabs}
\usepackage{array}
\usepackage{enumitem}
\usepackage{comment}
\usepackage{url}

\titleformat{\section}
  {\fontfamily{phv}\bfseries\fontsize{10}{12}\selectfont}
  {\thesection.}{0.5em}{}

\titleformat{\subsection}
  {\fontfamily{phv}\bfseries\fontsize{10}{12}\selectfont}
  {\thesubsection}{0.5em}{}

\newcommand{\ri}{r_i}
\newcommand{\ro}{r_o}
\newcommand{\Uh}{\hat{U}}
\newcommand{\uh}{\hat{u}}
\newcommand{\Th}{\hat{T}}
\newcommand{\ph}{\hat{p}}
\newcommand{\qh}{\hat{q}}
\newcommand{\Qh}{\hat{Q}}
\newcommand{\Ta}{\mathit{Ta}}
\renewcommand{\Pr}{\mathit{Pr}}
\newcommand{\Ro}{\Theta}           
\newcommand{\Lap}{\nabla^2}
\newcommand{\eps}{\varepsilon}
\newcommand{\sig}{\sigma}
\newcommand{\LBG}{\mathcal{L}_{\mathrm{BG}}}
\newcommand{\ddt}{\frac{\partial}{\partial t}}
\newcommand{\ddr}{\frac{\partial}{\partial r}}
\newcommand{\ddz}{\frac{\partial}{\partial z}}
\newcommand{\Ar}{\mathcal{A}_r}
\newcommand{\At}{\mathcal{A}_\theta}
\newcommand{\Az}{\mathcal{A}_z}

\begin{document}

\twocolumn[
\begin{center}

{\fontfamily{phv}\bfseries\fontsize{9}{11}\selectfont
Proceedings of the 1st International Conference on Innovations in Engineering
for Sustainable Transformations (InnovEST 2026)\\
April 3--4, 2026, NIT Jamshedpur, Jamshedpur-831014, Jharkhand, India
}

\vspace{0.3cm}

{\fontfamily{phv}\bfseries\fontsize{11}{13}\selectfont
InnovEST2026-236
}

\vspace{0.5cm}

{\fontfamily{phv}\bfseries\fontsize{12}{14}\selectfont
Linear Stability Analysis of convective flows in Rotating Baroclinic Annulus with Localized Peripheral Heating: A Floquet-BiGlobal Approach
}

\vspace{0.25cm}
{\fontfamily{phv}\fontsize{10}{12}\selectfont
\textbf{Jaya Nandan V}\textsuperscript{1},~\textbf{Ayan Kumar Banerjee}\textsuperscript{2*}}

\vspace{0.12cm}
{\fontfamily{phv}\fontsize{9.5}{11.5}\selectfont
\textsuperscript{1}Department of Physics, Amrita Vishwa Vidyapeetham, Coimbatore, Ettimadai, India\\
\textsuperscript{2}School of AI, Amrita Vishwa Vidyapeetham, Coimbatore, Ettimadai, India\\
*Corresponding Author: ayanbanerjee1@gmail.com}

\vspace{0.3cm}
Note: The following article has been submitted to Proceedings of the Innovations in Engineering for Sustainable Transformations 2026 and is Under Review. After it is published, it will be found at the Proceedings lecture notes. 
\vspace{0.3cm}

©2026 Jaya Nandan V and Ayan Kumar Banerjee. This article is distributed under a Creative Commons Attribution NonCommercial 4.0 International (CC BY-NC) License. https://creativecommons.org/licenses/by-nc/4.0/ 

\vspace{0.5cm}

\end{center}
]

{\fontfamily{phv}\bfseries ABSTRACT}

\noindent
We investigate the linear stability of a rotating fluid annulus subjected to localized heating at the outer periphery of the bottom surface and uniform cooling at the inner cylindrical wall through a rigorous stability analysis. The localized forcing generates a non-axisymmetric base state, invalidating the
classical normal-mode decomposition.  We employ Floquet--Bloch
theory in the azimuthal coordinate combined with a BiGlobal
eigenvalue formulation in the meridional ($r,z$) plane.  The
non-axisymmetric base state is expanded in azimuthal Fourier
harmonics; perturbations are expressed as quasi-periodic Bloch
modes that couple all azimuthal wavenumbers through base-state
harmonics.  Full linearised perturbation equations, the BiGlobal
block-operator structure, pressure elimination, solenoidal
projection, and the modal energy budget are derived.  Instability
is driven by cross-modal baroclinic energy release and shear
production---mechanisms absent in classical axisymmetric theory.

\vspace{0.3cm}

\section{INTRODUCTION}

The interplay between Coriolis and buoyancy forces in thermally
driven rotating flows underpins the dynamics of planetary
atmospheres, ocean currents, and stellar interiors.  A canonical
laboratory model is the rotating annulus: a fluid confined between
two coaxial cylinders rotating at angular velocity $\Omega$, with
the inner cylinder cooled and the outer wall heated, reproducing
baroclinically unstable jets analogous to mid-latitude atmospheric
circulation~\cite{Kaiser1971, Hide1965, Hide1977, Rossby1965}.

Although the rotating annulus configuration can reproduce flow structures reminiscent of the mid-latitude baroclinic zone in Earth’s atmosphere, a complete quantitative understanding of the mechanisms governing the formation of statically stable but baroclinically unstable regions remains elusive. This limitation stems from the classical baroclinic annulus, which employs isothermal walls and unidirectional forcing, thereby limiting the study of coupled baroclinic–stratification dynamics and inadequately representing atmospheric systems where bi-directional forcing—comprising both vertical and radial (meridional) thermal gradients—plays a dominant role~\cite{Banerjee2018}.

To overcome these limitations, Banerjee et al.~\cite{Banerjee2016,Banerjee2018,Banerjee2021} introduced a novel bi-directionally forced rotating convection experiment, consisting of a rotating fluid annulus subjected to uniform cooling at the inner cylindrical wall and localized heating at the outer base using a thin aluminium strip. This configuration enables the study of the interplay between baroclinic waves and the background stratification. Subsequent studies~\cite{banerjee2016iccms,banerjee2018thermacomp,Swarnakar2021,Banerjee2025a} reported the coexistence of columnar convective plumes adjacent to the outer boundary and baroclinic waves within the fluid interior, quantified the dependence of the Nusselt number ($Nu$) on the Taylor number ($Ta$) and heating rate. These studies revealed that $Nu$ is strongly influenced by buoyancy while remaining relatively insensitive to rotation. Further insights were obtained from two-dimensional axisymmetric simulations by Banerjee~\cite{Banerjee2024}, and more recent investigations into aspect-ratio effects~\cite{Banerjee2025b,Swarnakar2021,Banerjee2026} demonstrated that $Nu \sim Ra^{1/4}$ at moderate to high Rayleigh numbers, whereas at low $Ra$ and high $Ta$, rotational suppression significantly reduces heat transport. 

The localized
forcing produces an inherently non-axisymmetric base state, so
classical single-wavenumber stability analysis is inapplicable. The present paper derives a complete Floquet-BiGlobal stability
framework, providing the rigorous mathematical foundation for future
parametric studies in ($\Ta$, $\Ro$) space.

\section{GOVERNING EQUATIONS AND NON-DIMENSIONALISATION}

\subsection{Physical Configuration and Parameters}

An incompressible Boussinesq fluid occupies
$\ri\le r\le\ro$, $0\le\theta<2\pi$, $0\le z\le H$,
rigidly rotating at $\Omega\hat{z}$. Temperature boundary conditions:
inner wall ($r=\ri$): $T=T_i$ (cold); outer strip
($\ro-\delta_h\le r\le\ro$, $z=0$): $T=T_h$ (hot,
$\delta_h=5\,\mathrm{mm}$); all other boundaries thermally
insulating.

\begin{figure*}[!t]
\centering

\begin{minipage}{0.28\textwidth}
\centering
\includegraphics[width=\linewidth]{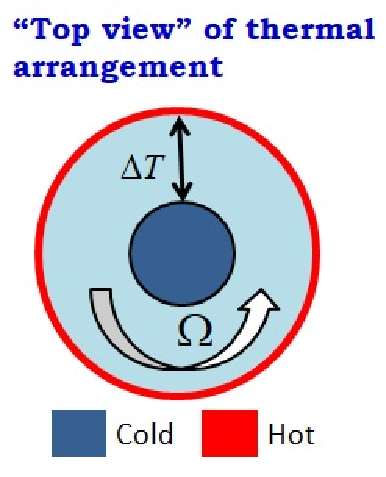}
\\ (a)
\end{minipage}
\hfill
\begin{minipage}{0.68\textwidth}
\centering
\includegraphics[width=\linewidth]{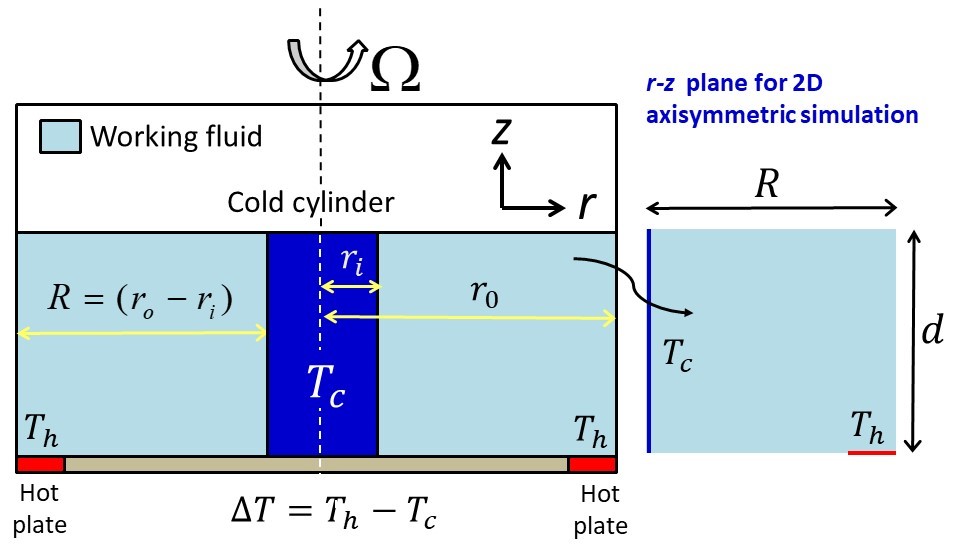}
\\ (b)
\end{minipage}

\caption{(a) Top view of the thermal forcing configuration. (b) Schematic of the rotating annulus with localized peripheral strip heating at the outer base.}
\label{fig:config}

\end{figure*}

\subsection{Non-Dimensionalisation}

Length scale $L=\ro-\ri$, velocity $U=\nu/L$, time $L/U$,
temperature $\Delta T=T_h-T_i$, pressure $\rho\nu^2/L^2$.
The governing dimensionless groups are:
\begin{align}
\Ta &= \frac{4\Omega^2 L^4}{\nu^2}
  \quad\text{(Taylor number)}
  \label{eq:Ta}\\
\Ro &= \frac{\beta g\Delta T\,L}{\Omega^2(\ri+\ro)}
  \quad\text{(thermal Rossby / Hide number)}
  \label{eq:Ro}\\
\Pr &= \frac{\nu}{\kappa}=7,
  \quad
  \eps=\frac{\delta_h}{L}\ll1
  \quad\text{(strip parameter)}
  \label{eq:Pr}
\end{align}

\subsection{Rotating-Frame Navier--Stokes System}

In the rotating frame, the dimensionless governing equations in
cylindrical coordinates $(r,\theta,z)$ with
$\mathbf{u}=(u_r,u_\theta,u_z)$ are:

\noindent\textit{Continuity:}
\begin{equation}
\frac{1}{r}\frac{\partial(ru_r)}{\partial r}
+\frac{1}{r}\frac{\partial u_\theta}{\partial\theta}
+\frac{\partial u_z}{\partial z}=0
\label{eq:cont}
\end{equation}

\noindent\textit{Radial momentum:}
\begin{align}
\ddt u_r+(\mathbf{u}\cdot\nabla)u_r
-\frac{u_\theta^2}{r}-2\Ta^{1/2}u_\theta
&=-\ddr p\notag\\
&\quad+\!\left(\Lap u_r-\frac{u_r}{r^2}
  -\frac{2}{r^2}\frac{\partial u_\theta}{\partial\theta}\right)
\label{eq:rmom}
\end{align}

\noindent\textit{Azimuthal momentum:}
\begin{align}
\ddt u_\theta+(\mathbf{u}\cdot\nabla)u_\theta
+\frac{u_r u_\theta}{r}+2\Ta^{1/2}u_r
&=-\frac{1}{r}\frac{\partial p}{\partial\theta}\notag\\
&\quad+\!\left(\Lap u_\theta-\frac{u_\theta}{r^2}
  +\frac{2}{r^2}\ddr u_r\right)
\label{eq:amom}
\end{align}

\noindent\textit{Axial momentum:}
\begin{equation}
\ddt u_z+(\mathbf{u}\cdot\nabla)u_z
=-\ddz p+\Lap u_z+\frac{\Ta\Ro}{4}\,T
\label{eq:zmom}
\end{equation}

\noindent\textit{Energy:}
\begin{equation}
\ddt T+(\mathbf{u}\cdot\nabla)T=\frac{1}{\Pr}\,\Lap T
\label{eq:energy}
\end{equation}

where $(\mathbf{u}\cdot\nabla)f=u_r\partial_r f
+(u_\theta/r)\partial_\theta f+u_z\partial_z f$ and
$\Lap f=(1/r)\partial_r(r\partial_r f)
+(1/r^2)\partial_{\theta\theta}f+\partial_{zz}f$.
The buoyancy factor $\Ta\Ro/4=\beta g\Delta T L^3/\nu^2$
is the modified Rayleigh number.

\section{NON-AXISYMMETRIC BASE STATE AND FOURIER DECOMPOSITION}

\subsection{Azimuthal Fourier Expansion}

Since the geometry is $2\pi$-periodic in $\theta$,
the steady base-state fields are expanded:
\begin{align}
\mathbf{U}_0(r,\theta,z)
  &=\sum_{n=-\infty}^{\infty}\Uh_n(r,z)\,e^{in\theta},
  \quad\Uh_{-n}=\Uh_n^*
  \label{eq:U0fourier}\\
T_0(r,\theta,z)
  &=\sum_{n=-\infty}^{\infty}\Th_n(r,z)\,e^{in\theta}
  \label{eq:T0fourier}
\end{align}

The $n=0$ component is the axisymmetric mean; $n\neq0$ modes
encode azimuthal structure forced by baroclinic nonlinearity.
Projecting the steady Navier--Stokes and energy system onto mode
$n$ (using the Fourier convolution theorem) gives the modal
continuity, energy, and axial-momentum equations with coupling
through convolution sums:
\begin{align}
\frac{1}{r}\ddr(r\Uh_{nr})+\frac{in}{r}\Uh_{n\theta}
+\ddz\Uh_{nz}&=0
\label{eq:mode_cont}\\
\sum_k\left[\Uh_k\cdot\nabla\Th_{n-k}\right]
  &=\frac{1}{\Pr}\,\mathcal{L}_n^T\Th_n
  \label{eq:mode_energy}\\
\sum_k\left[\Uh_k\cdot\nabla\Uh_{(n-k)z}\right]
  &=-\ddz\hat{P}_n+\mathcal{L}_n^z\Uh_{nz}
    +\frac{\Ta\Ro}{4}\Th_n
  \label{eq:mode_axial}
\end{align}

where the modal Laplacian is:
$\mathcal{L}_n^\alpha f=(1/r)\partial_r(r\partial_r f)
-(n^2+\eps_\alpha)/r^2\cdot f+\partial_{zz}f$,
with $\eps_\alpha=1$ for $r,\theta$-components and
$\eps_\alpha=0$ for $z,T$.  The convolution sums in
\eqref{eq:mode_energy}--\eqref{eq:mode_axial} couple all Fourier
harmonics, driving non-zero $\Uh_n$ ($n\neq0$) even when boundary
conditions are azimuthally uniform.

\subsection{Thermal Wind Balance}

In the rapid-rotation interior ($\Ta\gg1$), the dominant balance
relating each Fourier harmonic of the azimuthal jet to the
corresponding radial temperature gradient is:
\begin{equation}
2\Omega\,\frac{\partial\Uh_{n\theta}}{\partial z}
=-\beta g\Delta T\,\frac{\partial\Th_n}{\partial r}
\quad\text{for each }n
\label{eq:thermalwind}
\end{equation}

\section{LINEARISED PERTURBATION EQUATIONS}

\subsection{Reynolds Decomposition and Linearisation}

Decompose $\mathbf{u}=\mathbf{U}_0+\mathbf{u}'$
($|\mathbf{u}'|\ll|\mathbf{U}_0|$),
$T=T_0+T'$.  Substituting into
\eqref{eq:cont}--\eqref{eq:energy} and discarding
$O(|\mathbf{u}'|^2)$ terms:
\begin{align}
&\frac{1}{r}\ddr(ru_r')+\frac{1}{r}\frac{\partial u_\theta'}{\partial\theta}+\ddz u_z'=0
  \tag{L1}\label{eq:L1}\\
&\ddt u_r'+\Ar[\mathbf{U}_0,\mathbf{u}']
  -2\Ta^{1/2}u_\theta'-\frac{2U_{0\theta}u_\theta'}{r}
  \notag\\
&\quad=-\ddr p'
  +\!\left(\Lap u_r'-\frac{u_r'}{r^2}
    -\frac{2}{r^2}\frac{\partial u_\theta'}{\partial\theta}\right)
  \tag{L2}\label{eq:L2}\\
&\ddt u_\theta'+\At[\mathbf{U}_0,\mathbf{u}']
  +2\Ta^{1/2}u_r'
  +\frac{U_{0\theta}u_r'+U_{0r}u_\theta'}{r}
  \notag\\
&\quad=-\frac{1}{r}\frac{\partial p'}{\partial\theta}
  +\!\left(\Lap u_\theta'-\frac{u_\theta'}{r^2}
    +\frac{2}{r^2}\ddr u_r'\right)
  \tag{L3}\label{eq:L3}\\
&\ddt u_z'+\Az[\mathbf{U}_0,\mathbf{u}']
  =-\ddz p'+\Lap u_z'+\frac{\Ta\Ro}{4}\,T'
  \tag{L4}\label{eq:L4}\\
&\ddt T'+(\mathbf{U}_0\cdot\nabla)T'+(\mathbf{u}'\cdot\nabla)T_0
  =\frac{1}{\Pr}\,\Lap T'
  \tag{L5}\label{eq:L5}
\end{align}

\subsection{Linearised Advection Operators}

The complete linearised advection operators in
\eqref{eq:L2}--\eqref{eq:L4} are:
\begin{align}
\Ar[\mathbf{U}_0,\mathbf{u}']
&=U_{0r}\ddr u_r'
 +\frac{U_{0\theta}}{r}\frac{\partial u_r'}{\partial\theta}
 +U_{0z}\ddz u_r'
 \notag\\
&\quad+u_r'\ddr U_{0r}
 +\frac{u_\theta'}{r}\frac{\partial U_{0r}}{\partial\theta}
 +u_z'\ddz U_{0r}
 -\frac{2U_{0\theta}u_\theta'}{r}
\label{eq:Ar}\\
\At[\mathbf{U}_0,\mathbf{u}']
&=U_{0r}\ddr u_\theta'
 +\frac{U_{0\theta}}{r}\frac{\partial u_\theta'}{\partial\theta}
 +U_{0z}\ddz u_\theta'
 \notag\\
&\quad+u_r'\ddr U_{0\theta}
 +\frac{u_\theta'}{r}\frac{\partial U_{0\theta}}{\partial\theta}
 +u_z'\ddz U_{0\theta}
\label{eq:At}\\
\Az[\mathbf{U}_0,\mathbf{u}']
&=U_{0r}\ddr u_z'
 +\frac{U_{0\theta}}{r}\frac{\partial u_z'}{\partial\theta}
 +U_{0z}\ddz u_z'
 \notag\\
&\quad+u_r'\ddr U_{0z}
 +\frac{u_\theta'}{r}\frac{\partial U_{0z}}{\partial\theta}
 +u_z'\ddz U_{0z}
\label{eq:Az}
\end{align}

Key observation: equations \eqref{eq:L1}--\eqref{eq:L5} have
variable coefficients in \emph{all three} spatial directions through
$\mathbf{U}_0(r,\theta,z)$ and $T_0(r,\theta,z)$.  The
$\theta$-dependence of the base state precludes any classical
normal-mode reduction.

\section{FLOQUET-BLOCH THEORY IN THE AZIMUTHAL DIRECTION}

\subsection{Floquet-Bloch Theorem}

Since the linearised operator in \eqref{eq:L1}--\eqref{eq:L5} is
$2\pi$-periodic in $\theta$, the Floquet--Bloch theorem guarantees:
\begin{equation}
\mathbf{q}'(r,\theta,z,t)=e^{\sig t}\cdot e^{i\mu\theta}
\cdot\tilde{\mathbf{q}}(r,\theta,z),
\quad\tilde{\mathbf{q}}\text{ is }2\pi\text{-periodic}
\label{eq:floquet}
\end{equation}

Here $\sig=\sig_r+i\sig_i$ is the complex Floquet exponent
($\sig_r$ = growth rate, $\sig_i$ = oscillation frequency), and
$\mu\in[0,1)$ is the \textit{Bloch wavenumber} encoding azimuthal
quasi-periodicity.

\subsection{Fourier Expansion of the Bloch Mode}

Expanding the $2\pi$-periodic factor $\tilde{\mathbf{q}}$:
\begin{equation}
\tilde{\mathbf{q}}(r,\theta,z)
=\sum_{m=-\infty}^{\infty}\qh_m(r,z)\,e^{im\theta},
\quad
\qh_m=(\uh_{mr},\uh_{m\theta},\uh_{mz},\Th_m)^{\!\top}
\label{eq:bloch_fourier}
\end{equation}

so the full perturbation field is:
\begin{equation}
\mathbf{q}'=e^{\sig t}\sum_{m=-\infty}^{\infty}
\qh_m(r,z)\,e^{i(m+\mu)\theta}
\label{eq:fullpert}
\end{equation}

Each Fourier amplitude $\qh_m(r,z)$ is a 2-D field in the
meridional plane.  For $\mu=0$ and a single non-zero $m$,
\eqref{eq:fullpert} reduces to the classical normal mode.  The
Floquet framework couples \emph{all} modes $m\in\mathbb{Z}$ through
base-state harmonics.

\subsection{Coupled Floquet System (F1)--(F5)}

Substituting \eqref{eq:U0fourier}--\eqref{eq:T0fourier}
and \eqref{eq:floquet}--\eqref{eq:bloch_fourier} into
\eqref{eq:L1}--\eqref{eq:L5} and projecting onto azimuthal mode $m$:

\noindent\textit{Continuity} (F1):
\begin{equation}
\frac{1}{r}\ddr(r\uh_{mr})
+\frac{i(m+\mu)}{r}\uh_{m\theta}
+\ddz\uh_{mz}=0
\tag{F1}\label{eq:F1}
\end{equation}

\noindent\textit{Radial momentum} (F2):
\begin{align}
\sig\,\uh_{mr}
&+\sum_k\!\Bigl[
  \Uh_{kr}\ddr\uh_{(m-k)r}
  +\frac{i(k+\mu)\Uh_{k\theta}\uh_{(m-k)r}}{r}
  +\Uh_{kz}\ddz\uh_{(m-k)r}
  \notag\\
&\hphantom{+\sum_k\!\Bigl[}
  +\uh_{(m-k)r}\ddr\Uh_{kr}
  +\frac{i(k+\mu)\uh_{(m-k)\theta}\Uh_{kr}}{r}
  +\uh_{(m-k)z}\ddz\Uh_{kr}
  \notag\\
&\hphantom{+\sum_k\!\Bigl[}
  -\frac{2\Uh_{k\theta}\uh_{(m-k)\theta}}{r}
  -2\Ta^{1/2}\uh_{(m-k)\theta}\delta_{k0}
\Bigr]
  \notag\\
&=-\ddr\ph_m+\mathcal{L}_{(m+\mu)}^{\,r}\uh_{mr}
\tag{F2}\label{eq:F2}
\end{align}

\noindent\textit{Azimuthal momentum} (F3):
\begin{align}
\sig\,\uh_{m\theta}
&+\sum_k\!\Bigl[
  \Uh_{kr}\ddr\uh_{(m-k)\theta}
  +\frac{i(k+\mu)\Uh_{k\theta}\uh_{(m-k)\theta}}{r}
  +\Uh_{kz}\ddz\uh_{(m-k)\theta}
  \notag\\
&\hphantom{+\sum_k\!\Bigl[}
  +\uh_{(m-k)r}\ddr\Uh_{k\theta}
  +\frac{i(k+\mu)\uh_{(m-k)\theta}\Uh_{k\theta}}{r}
  +\uh_{(m-k)z}\ddz\Uh_{k\theta}
  \notag\\
&\hphantom{+\sum_k\!\Bigl[}
  +\frac{2\Uh_{k\theta}\uh_{(m-k)r}}{r}
  +2\Ta^{1/2}\uh_{(m-k)r}\delta_{k0}
\Bigr]
  \notag\\
&=-\frac{i(m+\mu)\ph_m}{r}+\mathcal{L}_{(m+\mu)}^{\,\theta}\uh_{m\theta}
\tag{F3}\label{eq:F3}
\end{align}

\noindent\textit{Axial momentum} (F4):
\begin{align}
\sig\,\uh_{mz}
&+\sum_k\!\Bigl[
  \Uh_{kr}\ddr\uh_{(m-k)z}
  +\frac{i(k+\mu)\Uh_{k\theta}\uh_{(m-k)z}}{r}
  +\Uh_{kz}\ddz\uh_{(m-k)z}
  \notag\\
&\hphantom{+\sum_k\!\Bigl[}
  +\uh_{(m-k)r}\ddr\Uh_{kz}
  +\uh_{(m-k)z}\ddz\Uh_{kz}
\Bigr]
  \notag\\
&=-\ddz\ph_m+\mathcal{L}_{(m+\mu)}^{\,z}\uh_{mz}+\frac{\Ta\Ro}{4}\,\Th_m
\tag{F4}\label{eq:F4}
\end{align}

\noindent\textit{Energy} (F5):
\begin{align}
\sig\,\Th_m
&+\sum_k\!\Bigl[
  \Uh_{kr}\ddr\Th_{m-k}
  +\frac{i(k+\mu)\Uh_{k\theta}\Th_{m-k}}{r}
  +\Uh_{kz}\ddz\Th_{m-k}
  \notag\\
&\hphantom{+\sum_k\!\Bigl[}
  +\uh_{(m-k)r}\ddr\Th_k
  +\frac{i(k+\mu)\uh_{(m-k)\theta}\Th_k}{r}
  +\uh_{(m-k)z}\ddz\Th_k
\Bigr]
  \notag\\
&=\frac{1}{\Pr}\,\mathcal{L}_{(m+\mu)}^{\,T}\Th_m
\tag{F5}\label{eq:F5}
\end{align}

The modal operators at quasi-wavenumber $(m+\mu)$ are:
\begin{equation}
\mathcal{L}_{(m+\mu)}^{\,\alpha} f
=\frac{1}{r}\ddr\!\left(r\ddr f\right)
-\frac{(m+\mu)^2+\eps_\alpha}{r^2}\,f
+\ddz^2 f
\label{eq:modal_lap}
\end{equation}

with $\eps_\alpha=1$ for $r,\theta$-components and
$\eps_\alpha=0$ for $z,T$.  The non-integer shift $\mu$ in
$(m+\mu)$ means that even integer azimuthal modes experience
non-integer effective wavenumbers---the fundamental Floquet
signature.

\section{BIGLOBAL EIGENVALUE PROBLEM}

\subsection{Global State Vector and Operator}

Truncating at $|m|\le M$ ($N_F=2M+1$ modes), the global state
vector is:
\begin{equation}
\Qh=(\qh_{-M},\,\qh_{-M+1},\,\ldots,\,\qh_0,\,\ldots,\,\qh_M)^\top
\label{eq:Qhat}
\end{equation}

The system \eqref{eq:F1}--\eqref{eq:F5} becomes the generalised
BiGlobal eigenvalue problem:
\begin{equation}
\sig\,\mathcal{M}\,\Qh=\LBG(\mu;\mathbf{U}_0,T_0)\,\Qh
\label{eq:EVP}
\end{equation}

where $\mathcal{M}$ is the mass matrix (identity for
velocity/temperature blocks; pressure handled via the
divergence-free constraint).  $\LBG$ is a block matrix of size
$(4N_F)\times(4N_F)$ acting in the meridional ($r,z$) plane:
\begin{equation}
\LBG=
\begin{pmatrix}
\mathcal{L}_{-M,-M} & \cdots & \mathcal{L}_{-M,M}\\
\vdots & \ddots & \vdots\\
\mathcal{L}_{M,-M} & \cdots & \mathcal{L}_{M,M}
\end{pmatrix}
\label{eq:LBG}
\end{equation}

\subsection{Off-Diagonal Coupling Blocks ($m\neq k$)}

Block $\mathcal{L}_{m,k}$ couples perturbation mode $k$ to mode
$m$ through base-state harmonic $(m-k)$:
\begin{align}
\mathcal{L}_{m,k}^{rr}&=
  -\Uh_{(m-k)r}\ddr
  -\frac{i(k+\mu)\Uh_{(m-k)\theta}}{r}
  -\Uh_{(m-k)z}\ddz
  \notag\\
&\quad-\ddr\Uh_{(m-k)r}
  +\frac{2\Uh_{(m-k)\theta}}{r}
\label{eq:Lmk_rr}\\
\mathcal{L}_{m,k}^{\theta r}&=
  -\ddr\Uh_{(m-k)\theta}
  -\frac{\Uh_{(m-k)\theta}}{r}
\label{eq:Lmk_tr}\\
\mathcal{L}_{m,k}^{zr}&=
  -\ddr\Uh_{(m-k)z}
\label{eq:Lmk_zr}\\
\mathcal{L}_{m,k}^{Tr}&=
  -\ddr\Th_{m-k}
  \quad\text{(thermal gradient coupling)}
\label{eq:Lmk_Tr}
\end{align}

\subsection{Diagonal Blocks ($m=k$) -- Viscous/Diffusive Operators}

The diagonal blocks carry the viscous/diffusive operators at
quasi-wavenumber $(m+\mu)$:
\begin{align}
\mathcal{L}_{m,m}^{rr}&=
  -\Uh_{0r}\ddr
  -\frac{i(m+\mu)\Uh_{0\theta}}{r}
  -\Uh_{0z}\ddz
  +\mathcal{L}_{(m+\mu)}^{\,r}
  \notag\\
&\quad-\ddr\Uh_{0r}
  +\frac{2\Uh_{0\theta}}{r}
  +2\Ta^{1/2}
\label{eq:Lmm_rr}\\
\mathcal{L}_{m,m}^{\theta\theta}&=
  -\Uh_{0r}\ddr
  -\frac{i(m+\mu)\Uh_{0\theta}}{r}
  -\Uh_{0z}\ddz
  +\mathcal{L}_{(m+\mu)}^{\,\theta}
  \notag\\
&\quad-\frac{\Uh_{0\theta}}{r}
  -2\Ta^{1/2}
\label{eq:Lmm_tt}\\
\mathcal{L}_{m,m}^{zz}&=
  -\Uh_{0r}\ddr
  -\frac{i(m+\mu)\Uh_{0\theta}}{r}
  -\Uh_{0z}\ddz
  +\mathcal{L}_{(m+\mu)}^{\,z}
\label{eq:Lmm_zz}\\
\mathcal{L}_{m,m}^{TT}&=
  -\Uh_{0r}\ddr
  -\frac{i(m+\mu)\Uh_{0\theta}}{r}
  -\Uh_{0z}\ddz
  +\frac{1}{\Pr}\mathcal{L}_{(m+\mu)}^{\,T}
\label{eq:Lmm_TT}
\end{align}

\subsection{Pressure Elimination and Solenoidal Projection}

Applying the divergence operator to the linearised momentum
equations and using \eqref{eq:F1} yields a Poisson equation for
each modal pressure:
\begin{equation}
\mathcal{L}_{(m+\mu)}^{\,2}\,\ph_m=R_m[\uh',\Th]
\label{eq:pressure_poisson}
\end{equation}

where $\mathcal{L}_{(m+\mu)}^{\,2}=(1/r)\partial_r(r\partial_r)
-(m+\mu)^2/r^2+\partial_{zz}$ and $R_m$ contains
base-state/perturbation interaction terms.  With Neumann condition
$\partial\ph_m/\partial n=0$ on all walls, \eqref{eq:pressure_poisson}
is uniquely invertible, yielding a pressure-free system in
$(\uh_{mr},\uh_{m\theta},\uh_{mz},\Th_m)$.

Alternatively, the toroidal-poloidal decomposition:
\begin{equation}
\uh_m=\nabla\times(\nabla\times\Phi_m\hat{z})
      +\nabla\times(\Psi_m\hat{z})
\label{eq:toro_polo}
\end{equation}

where $\Phi_m$ is the poloidal and $\Psi_m$ the toroidal scalar,
enforces continuity automatically and reduces degrees of freedom
from $4N_F$ to $3N_F$ per meridional grid point.

\subsection{Boundary Conditions on Modal Amplitudes}

\begin{align}
r=\ri:&\quad\uh_{mr}=\uh_{m\theta}=\uh_{mz}=0,\;
  \Th_m=0\quad\forall m
\tag{29a}\label{eq:bc_inner}\\
r=\ro:&\quad\uh_{mr}=\uh_{m\theta}=\uh_{mz}=0,\notag\\
&\quad\Th_m|_{z=0}=0\;\text{(strip)},\;
  \partial_n\Th_m=0\;\text{(elsewhere)}
\tag{29b}\label{eq:bc_outer}\\
z=0:&\quad\uh_{mr}=\uh_{m\theta}=\uh_{mz}=0,\notag\\
&\quad\Th_m=0\;\text{(strip)},\;
  \partial_z\Th_m=0\;\text{(elsewhere)}
\tag{29c}\label{eq:bc_bottom}\\
z=H:&\quad\partial_z\uh_{mr}=\partial_z\uh_{m\theta}
  =\uh_{mz}=0,\;\partial_z\Th_m=0
\tag{29d}\label{eq:bc_top}
\end{align}

\subsection{Numerical Discretisation}

The BiGlobal problem \eqref{eq:EVP} is discretised using
Chebyshev--Gauss--Lobatto (CGL) collocation: $N_r$ points in
$r\in[\ri,\ro]$ and $N_z$ points in $z\in[0,H]$.  Total matrix
size: $N_\mathrm{sys}=N_F\cdot N_r\cdot N_z\cdot 4$.  The
generalised EVP is solved via shift-invert Arnoldi
(ARPACK/SLEPc), targeting eigenvalues near shift $\sig_0$:
\begin{equation}
(\LBG-\sig_0\mathcal{M})^{-1}\mathcal{M}\,\Qh
=\frac{1}{\sig-\sig_0}\,\Qh
\label{eq:arpack}
\end{equation}

The analysis sweeps $\mu\in[0,1)$ to identify the most unstable
Bloch wavenumber $\mu^*$ and mode $m^*$ at marginal stability.

\section{ENERGY ANALYSIS AND INSTABILITY MECHANISMS}

\subsection{Total Perturbation Energy Budget}

Multiplying \eqref{eq:L2} by $u_r'^*$, \eqref{eq:L3} by
$u_\theta'^*$, \eqref{eq:L4} by $u_z'^*$, summing and integrating
over the fluid domain $V$, the kinetic energy
$E_k=\frac{1}{2}\int_V|\mathbf{u}'|^2\,dV$ evolves as:
\begin{align}
\frac{dE_k}{dt}&=P_S+P_B-D_k
\label{eq:Ek}\\
P_S&=-\int_V u_i'u_j'\frac{\partial U_{0i}}{\partial x_j}\,dV
\label{eq:PS}\\
P_B&=\int_V\frac{\Ta\Ro}{4}\,T'u_z'\,dV
\label{eq:PB}\\
D_k&=\int_V|\nabla\mathbf{u}'|^2\,dV
\label{eq:Dk}
\end{align}

The available potential energy
$E_p=\frac{2}{\Ta\Ro}\int_V T'^2\,dV$ satisfies:
\begin{align}
\frac{dE_p}{dt}&=-P_B-P_T-D_p
\label{eq:Ep}\\
P_T&=\frac{4}{\Ta\Ro}\int_V T'(\mathbf{u}'\cdot\nabla T_0)\,dV
\label{eq:PT}\\
D_p&=\frac{1}{\Pr}\frac{4}{\Ta\Ro}\int_V|\nabla T'|^2\,dV
\label{eq:Dp}
\end{align}

Adding \eqref{eq:Ek} and \eqref{eq:Ep}, total perturbation energy
$E=E_k+E_p$ obeys:
\begin{equation}
\frac{dE}{dt}=P_S+P_T-D_k-D_p
\label{eq:Etot}
\end{equation}

\textit{Instability criterion:}
$P_S+P_T>D_k+D_p$.  Production from shear extraction and
baroclinic tilting of isopycnals exceeds viscous and thermal
dissipation.

\subsection{Modal Energy Budget in Fourier Space}

Projecting \eqref{eq:Etot} onto Floquet mode $m$ using
\eqref{eq:fullpert}:
\begin{equation}
\sig_r\|\qh_m\|^2
=\sum_k\!\left[\hat{P}_S^{(m,k)}+\hat{P}_T^{(m,k)}\right]
-\hat{D}_k^m-\hat{D}_p^m
\label{eq:modal_budget}
\end{equation}

The cross-modal shear production coupling mode $m$ to all modes
$(m-k)$ through base-state harmonic $k$:
\begin{align}
\hat{P}_S^{(m,k)}&=
-\int\!\!\int\uh_{mr}^*\!\left[
  \uh_{(m-k)r}\ddr\Uh_{kr}
  +\frac{i(k+\mu)\uh_{(m-k)\theta}\Uh_{kr}}{r}
  +\uh_{(m-k)z}\ddz\Uh_{kr}+\cdots
\right]r\,dr\,dz+\mathrm{c.c.}
\label{eq:PS_modal}
\end{align}

Equation~\eqref{eq:PS_modal} shows that perturbation mode $m$ is
energized by base-state harmonic $k$ acting on perturbation mode
$(m-k)$---a \textit{cross-modal energy transfer} that is the
energetic signature of Floquet coupling, entirely absent in
classical axisymmetric stability theory.

\subsection{Instability Types Identified by Energy Budget}

\noindent\textbf{Baroclinic:} $P_T$ dominates.  Potential energy
of inclined isopycnals (sustained by radial temperature gradient)
converts to kinetic energy.  Characteristic of moderate $\Ta$,
intermediate $\Ro$.

\noindent\textbf{Barotropic:} $P_S$ dominates ($k=0$ term).
Shear of azimuthal jet $\partial\Uh_{0\theta}/\partial r$ extracts
mean-flow energy, enhanced by non-axisymmetric jets generated by
the strip.

\noindent\textbf{Parametric (Floquet resonance):}
Cross-modal $\hat{P}_S^{(m,k)}$ with $k\neq0$.  Occurs when
rational $\mu=p/q$ causes modes $m$ and $m+q$ to become degenerate
through base-state harmonic $q$.

\section{ASYMPTOTIC LIMITS AND PHYSICAL INTERPRETATION}

\subsection{Axisymmetric Limit ($\eps\to0$)}

When $\delta_h\to0$, $\Uh_n\to0$ for $n\neq0$.  All off-diagonal
blocks $\mathcal{L}_{m,k}$ ($m\neq k$) vanish, and \eqref{eq:EVP}
decouples into independent classical normal-mode problems:
\begin{equation}
\sig\,\qh_m=\mathcal{L}_{m,m}^{(0)}\,\qh_m
\quad\forall m\in\mathbb{Z}
\quad(\text{classical result})
\label{eq:classical}
\end{equation}

confirming that the BiGlobal framework is a strict generalization,
recovering classical theory for $\eps\to0$.

\subsection{Eady-Like Criterion and Floquet Resonances}

Near baroclinic onset, with quasi-axisymmetric base state, the
critical mode $m_c$ satisfies the modified Eady criterion:
\begin{equation}
m_c\approx\frac{2\pi\bar{r}}{L_R},
\quad
L_R=\frac{NH}{\Omega},
\quad
N=\sqrt{\frac{\beta g\Delta T}{H}}
\label{eq:eady}
\end{equation}

where $\bar{r}=(\ri+\ro)/2=0.135\,\mathrm{m}$.  The strip reduces
the local $L_R$ near $r=\ro$, preferentially exciting higher-$m$
modes.  At rational $\mu=p/q$, Floquet resonance occurs when modes
$m$ and $m+q$ become degenerate through the base-state harmonic
$\Uh_q(r,z)$:
\begin{equation}
\mathcal{L}_{m+q,m}\sim\Uh_q(r,z)\neq0
\quad\text{when }\eps>0
\label{eq:resonance}
\end{equation}

This parametric instability appears as narrow tongues of enhanced
growth rate in $(\Ta,\Ro)$ space---the rotating-annulus analog of
Mathieu-equation resonance bands.

\section{CONCLUSIONS}

A self-contained Floquet–BiGlobal linear stability framework has been developed for the rotating baroclinic annulus with localized peripheral heating. The non-axisymmetric base state is represented through an expansion in azimuthal Fourier harmonics, wherein the steady governing equations couple adjacent modes via convolution terms arising from baroclinic nonlinearities. Perturbations are introduced using a Floquet–Bloch ansatz of the form $e^{\sigma t + i\mu\theta}\tilde{\mathbf{q}}(r,\theta,z)$, which enables simultaneous coupling of all azimuthal modes through the Floquet parameter $\mu \in [0,1)$. The resulting fully coupled Floquet system explicitly incorporates Coriolis, curvature, and thermal-gradient effects associated with each base-state harmonic. This leads to a BiGlobal operator structure in which diagonal blocks contain viscous and diffusive operators evaluated at non-integer quasi-wavenumbers $(m+\mu)$, while off-diagonal blocks capture inter-modal interactions mediated by the base state. A pressure-free, divergence-free formulation is obtained through pressure elimination and solenoidal projection, allowing efficient numerical solution via shift-invert Arnoldi methods. An analysis of the modal energy budget reveals that cross-modal shear production terms, particularly those involving interactions between different harmonics ($k \neq 0$), serve as the primary energetic signature of Floquet coupling—a mechanism absent in classical axisymmetric stability theory. In the asymptotic limit of vanishing non-axisymmetry, the formulation recovers the classical results, while for finite amplitudes, Floquet resonances at rational values of $\mu = p/q$ give rise to parametric instability tongues in the $(\Ta,\Ro)$ parameter space.



\begin{thebibliography}{7}

\bibitem{lorenz1963}
E.~N.~Lorenz, ``Deterministic nonperiodic flow,'' \textit{J.~Atmos.~Sci.}, vol.~20, pp.~130--141, 1963.

\bibitem{ghil1987}
M.~Ghil and S.~Childress, \textit{Topics in Geophysical Fluid Dynamics}, Springer, New York, 1987.

\bibitem{Banerjee2024}
A.K.~Banerjee, Axisymmetric Study of Convection in Rotating
Annulus in the Presence of Localized Heating,
\textit{Physics of Fluids}, Vol.~36, Issue~12, 126621,
December 2024. doi:~\url{https://doi.org/10.1063/5.0239746}.

\bibitem{Banerjee2021}
A.K.~Banerjee, A.~Bhattacharya, S.~Balasubramanian,
Investigation of Heat Transfer Characteristics in a Rotating
Convection System with Bidirectional Thermal Gradients,
\textit{ASME J.\ Heat Transfer}, January 2021; 143(1):011802.
doi:~\url{https://doi.org/10.1115/1.4048825}.

\bibitem{Banerjee2018}
A.K.~Banerjee, A.~Bhattacharya, S.~Balasubramanian,
Experimental Study of Rotating Convection in the Presence of
Bi-directional Thermal Gradients with Localized Forcing,
\textit{AIP Advances}, Vol.~8, Issue~11, 115324, 2018.
doi:~\url{https://doi.org/10.1063/1.5061808}.

\bibitem{Swarnakar2021}
S.~Swarnakar, A.K.~Banerjee, A.~Bhattacharya,
S.~Balasubramanian, Numerical Investigation of Rotating
Convection in a New Configuration with Bidirectional Thermal
Gradients, in: T.~Prabu, P.~Viswanathan, A.~Agrawal,
J.~Banerjee (eds.), \textit{Fluid Mechanics and Fluid Power},
Lecture Notes in Mechanical Engineering, Springer, Singapore,
2021. doi:~\url{https://doi.org/10.1007/978-981-16-0698-4_56}.

\bibitem{Kaiser1971}
J.~Kaiser, Heat Transfer by Symmetrical Rotating Annulus Convection,
\textit{Journal of the Atmospheric Sciences}, Vol.~28, pp.~929--932, 1971. doi:~\url{https://doi.org/10.1175/1520-0469(1971)028<0929:HTBSRA>2.0.CO;2}.


\bibitem{Hide1965}
R.~Hide and W.~Fowlis, Thermal Convection in a Rotating Annulus of Liquid: Effect of Viscosity on the Transition Between Axisymmetric and Non-Axisymmetric Flow Regimes,
\textit{Journal of the Atmospheric Sciences}, Vol.~22, pp.~541--558, 1965. doi:~\url{https://doi.org/10.1175/1520-0469(1965)022<0541:TCIARA>2.0.CO;2}.


\bibitem{Hide1977}
R.~Hide, P.~Mason, and R.~Plumb, Thermal Convection in a Rotating Fluid Subject to a Horizontal Temperature Gradient: Spatial and Temporal Characteristics of Fully Developed Baroclinic Waves,
\textit{Journal of the Atmospheric Sciences}, Vol.~34, pp.~930--950, 1977. doi:~\url{https://doi.org/10.1175/1520-0469(1977)034<0930:TCIARF>2.0.CO;2}.

\bibitem{Banerjee2025a}
A.~K.~Banerjee and S.~Swarnakar, ``Aspect ratio dependence in the convection in rotating annulus in the presence of localized heating,'' in \textit{Recent Advances in Thermal and Fluid Science}, A.~K.~Parwani, D.~K.~Gupta, V.~M.~Patel, and S.~Ray, Eds., Lecture Notes in Mechanical Engineering, Springer, Singapore, 2026, doi: 10.1007/978-981-96-8508-0\_1.

\bibitem{Banerjee2025b}
A.~K.~Banerjee, ``An integrated laboratory and axisymmetric numerical study of convection in a rotating annulus with bi-directional thermal forcings,'' in \textit{Recent Advances in Thermal and Fluid Science}, A.~K.~Parwani, D.~K.~Gupta, V.~M.~Patel, and S.~Ray, Eds., Lecture Notes in Mechanical Engineering, Springer, Singapore, 2026, doi: 10.1007/978-981-96-8508-0\_3.

\bibitem{Banerjee2016}
A.K.~Banerjee, S.~Tirodkar, A.~Bhattacharya,
S.~Balasubramanian, Convection in Rotating Flows with
Simultaneous Imposition of Radial and Vertical Temperature
Gradients, \textit{VIII$^{\text{th}}$ Intl.\ Symp.\ on
Stratified Flows}, August 29--September~1, 2016, San Diego,
CA, USA. arXiv:1611.00807.


\bibitem{Rossby1965}
H.T.~Rossby, On thermal convection driven by non-uniform heating from below: an experimental study,
\textit{Deep Sea Research and Oceanographic Abstracts}, Vol.~12, pp.~9--16, 1965. doi:~\url{https://doi.org/10.1016/0011-7471(65)91336-7}.

\bibitem{banerjee2018thermacomp}
A.~K.~Banerjee, A.~Bhattacharya, and S.~Balasubramanian, ``Experimental study of rotating convection in a novel configuration,'' in \textit{Proc.~5th Int.~Conf.~on Computational Methods for Thermal Problems (ThermaComp2018)}, Bangalore, India, Jul.~9--11, 2018, ISSN 2305-6924.

\bibitem{banerjee2016iccms}
A.~K.~Banerjee, A.~Bhattacharya, and S.~Balasubramanian, ``Effect of rotation and baroclinicity on heat transport and turbulent convection in annular flow,'' in \textit{Proc.~6th Int.~Congress on Computational Mechanics and Simulation (ICCMS)}, IIT Bombay, India, Jun.~27--Jul.~1, 2016.

\bibitem{Banerjee2026}
A.K.~Banerjee, S.~Swarnakar, Numerical Influence of Aspect ratio in the Convection in Rotating Annulus In the Presence of Localized Heating, \textit{FMFP
Lecture Notes in Mechanical Engineering}, Springer Singapore,
2026 (under review)

\end{thebibliography}
\end{document}